\documentclass[12pt]{iopart}
\usepackage{bbm, dsfont}
\usepackage{graphicx, graphics}
\usepackage{iopams}
\usepackage{float}
\expandafter\let\csname equation*\endcsname\relax
\expandafter\let\csname endequation*\endcsname\relax
\usepackage{amsmath,amsfonts,amsthm}
\usepackage{tikz-cd} 

\usepackage{hyperref}

\begin{document}
\title[Machine learning order paramters dynamics]{Machine learning stochastic differential equations for the evolution of order parameters of classical many-body systems in and out of equilibrium\\
}

\author{Francesco Carnazza$^1$, 
        Federico Carollo$^1$,
        Sabine Andergassen$^2$,
        Georg Martius$^{3,4}$ and
        Miriam Klopotek$^5$,
        Igor Lesanovsky$^{1,6}$}
\address{$^1$ Institut für Theoretische Physik and Center for Quantum Science,
        Universität Tübingen, 
        Auf der Morgenstelle 14, 
        72076 Tübingen, Germany}
\address{$^2$ Institute of Information Systems Engineering and Institute for Solid State Physics, Vienna University of Technology, 1040 Vienna, Austria}

\address{$^3$ Max Planck Institute for Intelligent Systems,
              Max-Planck-Ring 4, 72076 Tübingen, Germany}
\address{$^4$ Wilhelm Schickard Institut für
Informatik, Universit\"{a}t T\"{u}bingen, Maria-von-Linden-Straße 6,
72076 T\"{u}bingen,
Germany}
\address{$^5$ Stuttgart Center for Simulation Science, SimTech Cluster of Excellence EXC 2075, University of Stuttgart, Universitätsstraße 32,
 70569 Stuttgart,
 Germany}
\address{$^6$ Centre for the Mathematics and Theoretical 
            Physics of Quantum Non-Equilibrium Systems,
            University of Nottingham, Nottingham, NG7 2RD, UK}

\begin{abstract}
We develop a machine learning algorithm to infer the emergent stochastic equation governing the evolution of an order parameter of a many-body system. 
We train our neural network to independently learn the directed force acting on the order parameter as well as an effective diffusive noise. We illustrate our approach using the classical Ising model endowed with Glauber dynamics, and the contact process as test cases.  
For both models, which represent paradigmatic equilibrium and nonequilibrium scenarios, the directed force and noise can be efficiently inferred. The directed force term of the Ising model allows us to reconstruct an effective potential for the order parameter which develops the characteristic double-well shape below the critical temperature. Despite its genuine nonequilibrium nature, such an effective potential can also be obtained  for the contact process and its shape signals a phase transition into an absorbing state. Also, in contrast to the equilibrium Ising model, the presence of an absorbing state renders the noise term dependent on the value of the order parameter itself.
\end{abstract}
\section{Introduction}
\label{sec:intro}
\begin{figure*}
        \centering
        \includegraphics[width=1.\linewidth]{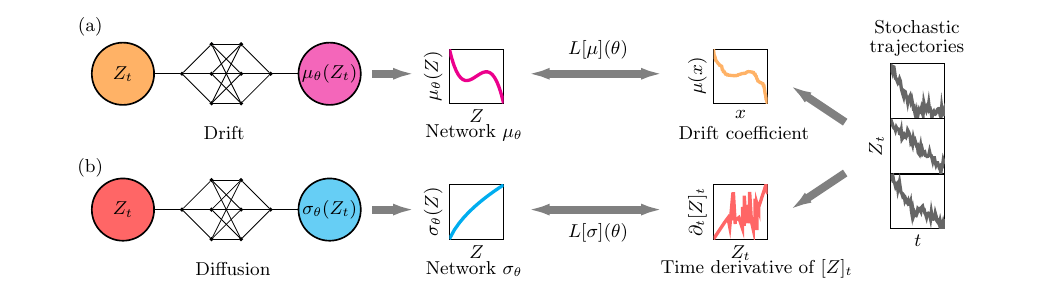}
        \caption{
        \textbf{Training of the drift coefficient $\mu_\theta$ and the diffusion coefficient  $\sigma_\theta$ from the stochastic trajectories of the order parameter.} 
        (a) The network for the drift coefficient, $\mu_\theta$, is trained 
        by minimizing a distance, $L[\mu](\theta)$ in Eq.~(\ref{eq:loss_drift}), between $\mu_\theta(x)$ and
        the drift coefficient $\mu$, cf.~Eq.~(\ref{eq:inf_gen}).
        For a given value $x$ of the order parameter, we consider all the values $Z_t$ close to it in the observed trajectories, and their update $Z_{t+dt}$.
        The finite difference $Z_{t+dt}-Z_{t}$ is then used to locally approximate the drift coefficient $\mu$.
        (b) The network for the diffusion coefficient $\sigma_\theta$ is trained by minimizing a distance, cf. Eq. (\ref{eq:loss_diffusion}), between $\sigma^2_\theta(Z_t)$ and the derivative of the quadratic variation $\partial_t[Z]_t$, see $L[\sigma](\theta)$ in Eq.~(\ref{eq:d_quad_var}).
        The quadratic variation $[Z]_t$ itself, cf. Eq. (\ref{eq:quad_var}), is locally approximated from the observed trajectories.}
        \label{fig:figure0}
\end{figure*}
Stochastic processes are fundamentally important in physics \cite{Kampen2007, Coffey1996, Gillespie1992}. For instance, random microscopic fluctuations can strongly impact the evolution of macroscopic physical observables, e.g., order parameters close to phase transitions.
Monte Carlo methods \cite{Metropolis1953,Hohenberg1964,Rosenbluth2004} are often the ``benchmark" for the computational treatment of classical many-body dynamics, allowing for efficient sampling of stochastic microscopic configurations and trajectories.
The large-scale dynamics of the order parameter are instead typically modeled by a stochastic differential equation. The latter contains both a force term, leading to a deterministic drift, and a noise term yielding diffusive behavior 
\footnote{The drift and the diffusion represent the most basic ingredients for a coarse-grained dynamics. More general forms might include memory kernels or other non-Markovian time dependencies \cite{Widder2022, Schilling20221}.
}.
\vphantom{\cite{Widder2022, Schilling20221}}
However, establishing a connection between fluctuating microscopic stochastic trajectories and the coarse-grained evolution of the order parameter is a challenging task that can rarely be accomplished analytically.

In this paper, we develop a machine learning approach \cite{Carleo2019rev, Mehta_2019, grogan2020} to bridge this gap. To illustrate our method, we consider two paradigmatic classical many-body systems: the 2D Ising model evolving under Glauber dynamics \cite{Glauber2004,Mehmet2014,Walter2015} and the nonequilibrium contact process in 1D. The dynamics considered for the Ising model obey detailed balance, which eventually takes the system to a state of thermal equilibrium. As a function of temperature, this state shows a transition from a paramagnetic to a ferromagnetic state, characterized by a zero and non-zero value of the order parameter, respectively. As we will show, this transition manifests in the structure of the learned drift term cf.~Fig.~\ref{fig:figure0}(a), from which one can reconstruct an effective potential that exhibits a characteristic double-well shape  below the critical temperature. Both the paramagnetic and ferromagnetic phases are fluctuating, which is also reflected in the learned noise term.
In contrast to the scenario of the  Ising model, the contact process represents a genuine out-of-equilibrium system \cite{Majumdar1999, Callen1985, Halpin1998,Halpin1991}, i.e., its dynamics does not obey detailed balance and its stationary state is non-thermal. The model features a phase transition between a non-fluctuating absorbing state in which the order parameter is strictly zero and a fluctuating active phase with a non-vanishing order parameter. Interestingly, we show that also for this genuine nonequilibrium process, an effective potential governing the deterministic drift of the order parameter can be constructed using our machine learning approach. Unlike for the Ising model, however, where the learned noise is such that both phases are fluctuating, a noise term is inferred whose strength depends on the order parameter. In particular, the noise strength tends to zero for vanishing values of the order parameter, see sketch in Fig.~\ref{fig:figure0}(b), signalling an approach to the (non-fluctuating) absorbing state.

Our method is applicable to a wide range of many-body processes in and out of equilibrium. It provides a way to determine a stochastic equation for order parameters which is intuitive and directly interpretable, as in mean-field theories. Remarkably, it also carries information about the exact low-dimensional physics of the considered model, as we demonstrate through estimates of critical exponents.  Moreover, our method  should also be applicable for inferring effective stochastic differential equations for the evolution of order parameters from experimental data. 
In contrast with other  machine learning routines, which learn the stochastic differential equation by integrating  
the stochastic dynamics 
and optimizing over the probability distribution of the  variable,
our approach builds on learning ordinary differential equations \cite{MorrillSKF21, Kidger2020a,Chen2018,Toth2020Hamiltonian, Goodfellow2014, kidger2021,Dietrich2022}.  
\section{Many-body stochastic processes}
\label{sec:stocproc}
\subsection{The evolution of stochastic observables}
For the sake of concreteness, we focus on many-body lattice systems %made 
of $N$ sites, each of which is associated with a   classical spin variable. We denote the system state, or system configuration, through the vector $s$ containing the values $s^i$ of the variables at the different sites $i$.
We furthermore assume the system to be subject to a discrete-time Markovian stochastic spin-flip dynamics. 

Relevant information about the above many-body system is  provided by so-called order parameters, which encode properties of the whole configuration. A paradigmatic example is given by an average of the form  $Z_t \equiv Z(s_t) =  \frac{1}{N}\sum_{i} s_t^i$, where $s_t$ is the time-evolving state of the system. As a consequence of the stochastic nature of $s_t$, also the effective dynamics of $Z_t$ is stochastic. For large systems and at a continuous coarse-grained time scale,  $Z_t$ becomes a continuous random variable that may be expected to obey an emergent  stochastic differential equation of the form 
\begin{equation}
    dZ_t = \mu(Z_t,t) dt + \sigma(Z_t,t)  dW_t\,. 
    \label{eq:sde_ito}
\end{equation}
Here, the function $\mu$ is referred to as the drift coefficient, 
while 
$\sigma$ 
is called  diffusion coefficient. $W_t$ is a standard Wiener process \cite{Kampen2007} and $dW_t$ is its increment satisfying the relations $\mathbb{E}[dW_t]=0$ and $\mathbb{E}[dW_t^2]=dt$, with $\mathbb{E}$ denoting expectation over the noise. Despite the simple form of Eq.~\eqref{eq:sde_ito}, understanding the functional form of  $\mu$ and $\sigma$ is in general a difficult task.
In the following, we propose a method to learn an approximation to the analytical form of the drift and the diffusion coefficients by means of neural networks.
We determine 
two artificial neural networks $\mu_\theta$ and $\sigma_\theta$ (see sketch in Fig.~\ref{fig:figure0}), which describe the dynamics of $Z_t$, given the network parameters (weights and biases) $\theta$.
We restrict ourselves to the Markovian case in which $\mu_\theta$ and $\sigma_\theta$ do not depend on time
\begin{equation}
    dZ_t = \mu_\theta(Z_t) dt + \sigma_\theta(Z_t)  dW_t.
    \label{eq:nn_sde_ito}
\end{equation}
To approximate the functions $\mu$ and $\sigma$ we use a data-driven method, i.e., the networks $\mu_\theta$ and $\sigma_\theta$ are trained
on a data set composed of trajectories $Z_t$, which we call ground truth data, see also Fig.~\ref{fig:figure0}. 
Note that restricting to the Markovian case of Eq.~\eqref{eq:nn_sde_ito} is an assumption since, even if the dynamics of the system configuration $s_t$ is Markovian at the microscopic scale, the emergent dynamics of the order parameters -- i.e. macroscopic quantities -- may feature non-Markovian effects.

\subsection{Neural network representation of the drift and diffusion coefficients}
\label{sec:drift_diffusion}
Our approach consists of training the  networks $\mu_\theta$ and $\sigma_\theta$ with separate routines and not simultaneously. As we discuss below, this allows us to train them on average quantities.
The drift term can be quantified by exploiting averages over trajectories of the infinitesimal increment $dZ_t$ in Eq.~\eqref{eq:sde_ito}. More precisely, starting from Eq.~\eqref{eq:sde_ito} it is possible to show that the function $\mu$ at point $x$ can be obtained as the limit \cite{pavliotis2016stochastic, Dynkin965,Oksendal2014}
\begin{equation}
\mu(x)
=
    \lim _{dt \rightarrow 0^+}
    \frac{
        \mathbbm{E}_{Z_t=x}[Z_{t+dt}]-x
    }{
        dt
    }
    ,
    \label{eq:inf_gen}
\end{equation}
where $\mathbbm{E}_{Z_t=x}$ denotes expectation conditional on the process being in $x$ at time $t$. In the theory of stochastic processes, the above limit also provides the action of the so-called infinitesimal generator $\mathcal{W}$ on the function $x$, $\mu(x)=\mathcal{W}[x]$ \cite{Oksendal2014}. 

\begin{figure}
    \centering	
    \includegraphics[width=0.7\textwidth]{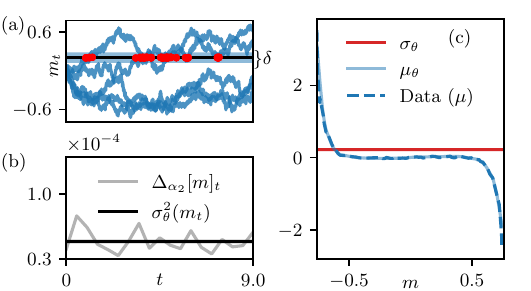}
    \caption{
    \textbf{
    Estimation of the drift and diffusion coefficients $\mu$ and $\sigma$.} 
    In panel (a), some exemplary trajectories $Z_t$ for the Ising model on a $128 \times 128$ square lattice are depicted. Here, the order parameter $Z_t$ is the magnetization  $m_t$. These trajectories are generated using the dynamical rules reported in Eq.~(\ref{eq:glauberMH}) and provide the variable $m_t$.
    For a given $m_t$ (black solid line), all the $n_i$ intersections (red dots) within the shaded area 
    are used to compute the quantity 
    $\mu(x)dt$.
    The latter is plotted in panel (c) (dashed line), while the solid line is obtained from the trained network $\mu_\theta(m)$.
    In panel (c) the network for the diffusion coefficient $\sigma_\theta(m)$ is also shown, in order to compare its dependence on the magnetization $m$ with the one of the drift.
    This can also be seen in panel (b), where the values for the finite difference
    of the quadratic variation $\Delta_{\alpha_2} [m]_t$,
    upon which the network is trained, are provided together with the values of the corresponding
    predictions of the neural network $\sigma_\theta^2(m_t)$. 
}
    \label{fig:mu_loss_ising}
\end{figure}

The limit in Eq.~\eqref{eq:inf_gen} can be estimated from the data set, as sketched in Figs.~\ref{fig:figure0}-\ref{fig:mu_loss_ising}. 
To this end, we generate batches 
$X_i=\{x_1,...,x_{d_{\rm batch}}\}$ of size $d_{\rm batch}$.
Each $x_j $ in $ X_i$ is extracted randomly between the minimum and maximum values of the trajectories $Z_t$.
For each $x_j$, we consider all the $n_j$ points $Z_t^j$, in all trajectories,  which belong to the interval of width  $\delta$ around $x_j$, see Fig.~\ref{fig:mu_loss_ising}(a). 
The value of $\delta$ has to be chosen in such a way that all bins associated with the different $x_j$ are sufficiently populated, ensuring the smoothness of the learned $\mu(x_j)$. 
We check \textit{a posteriori} that the predicted dynamics, learned with such a $\delta$, corresponds to the ground truth (see \ref{nn_ad_int}).
\footnote{
This interval is defined as 
$B^\delta_j \equiv 
\{ \text{All } Z_t \text{ such that } |Z_t-x_j|<\delta \}
$, and its cardinality is $n_j \equiv \# B_\delta^j$. 
We denote the points in this interval as $Z_t^j$. 
}. 
We optimize $\mu_\theta$ by minimizing the following loss function, cf. Fig. \ref{fig:mu_loss_ising}(b)
\begin{equation}
    L[\mu_\theta](\theta) = 
    \sum_{j = 1}^{d_{\rm batch}} 
    \left| 
    \mu_\theta(x_j)
    -
    \frac{1}{n_j} 
    \sum _{Z_t^j}  
    \Delta_1 Z_t^j
    \right|,\label{eq:loss_drift}
\end{equation}
where $\Delta_1 Z_t \equiv (Z_{t+d t } - Z_{t})/d t $. 
We consider the coarse-grained adimensional time $t$ to correspond to the number of discrete-time updates of the system normalized by a suitable factor $\tau$ and thus  $dt=1/\tau$.

In our data sets, the observed noise is often larger than the drift, cf.~Fig.~\ref{fig:mu_loss_ising}(c)-\ref{fig:sigma_cp}(c), especially near the stationary state, where the drift coefficient vanishes altogether. 
This is why computing the targets 
$\frac{1}{n_j} 
\sum _{Z_t^j}  
\Delta_1 Z_t^j$ 
in Eq.~\eqref{eq:loss_drift} is essential.
In fact, no learning would be possible without taking the targets to be arithmetic averages, due to the above-mentioned large fluctuations. 

Since our task is to understand the order-parameter dynamics, we restrict ourselves to the problem of learning one-dimensional data.
This allows for an efficient estimate of the drift coefficient in Eq.~\eqref{eq:nn_sde_ito}. In one dimension, the stochastic quantity $Z_t$
indeed hits the different intervals sufficiently many times during the evolution, which is needed for proper sampling and computing $\mu(x)$.
To reduce over-fitting, we train $n_{\rm train}$ different models $\mu_\theta^i(x)$, 
with the loss function~\eqref{eq:loss_drift}.
To each of these models, we assign a weight $w_i$ equal to the inverse of the mean square error between the data estimate of $\mu(x)$ 
and the network
result  $\mu_\theta^i$.
As a reference model $\mu_\theta$, we take the weighted average over this ``ensemble" of models:
\begin{equation}
    \mu_\theta = \sum_{i=1}^{n_{\rm train}} \frac{w_i \mu_\theta^i}{ \sum_{i=j}^{n_{\rm train}}w_j}.
\end{equation}
The values of $n_{\rm train}$, $d_{\rm batch}, n_{\rm epochs}$ and $\delta$ we adopt for the considered models are reported in  Table \ref{fig:net_details} (see \ref{nn_ad_int}).

In order to learn the diffusion coefficient, we use the ``second moment" of $dZ_t$, which is the quadratic variation $[Z]_t$. For stochastic processes as in Eq.~\eqref{eq:sde_ito}, this is given by \cite{Protter1990, Karandikar2014}
\begin{equation}
    [Z]_t = \int_0^t  dZ_s^2=\int_0^t ds \sigma^2 (Z_s)\, ,
    \label{eq:quad_var}
\end{equation}
which is nothing but the integral version of the  
differential equation
\begin{equation}
    \partial_t [Z]_t = \sigma ^2 (Z_t).
    \label{eq:qv_der}
\end{equation}
To train the network for the diffusion coefficient $\sigma_\theta$, we  devise a coarse-graining procedure that makes the spin-flip noise of the stochastic many-body dynamics look like a Wiener process. To this end, we first compute  the quadratic variation from trajectories as
\begin{equation}
[Z]_t\cong \sum_{u\in[0,t]} \left(Z_{u+\alpha_1 du}-Z_u\right)^2\, .
    \label{eq:Z_var_cg}
\end{equation}
Here, the integer factor $\alpha_1\ge1$ may allow one to magnify the variation at the different times. 
Furthermore, we approximate  Eq.~\eqref{eq:qv_der} by 
\begin{equation}
    \partial_t [Z]_t \approx
    \Delta_{\alpha_2} [Z]_t =  \frac{[Z]_{t+\alpha_2 d t } - [Z]_{t}}{ \alpha_2  dt }.
    \label{eq:d_quad_var}
\end{equation}

The factor $\alpha_2\gg1$ allows one to coarse-grain the noise over many discrete time-steps, which  proved necessary for convergence during the training procedure. 
This is mainly due to the fact that the finite difference in Eq.~\eqref{eq:d_quad_var} is stochastic. For this reason we need an average in order to obtain valuable information for the training. %The choice $\Delta t \gg dt$ is equivalent  to an average over time.
Eq.~\eqref{eq:d_quad_var} will still be a good approximation of a time derivative if we consider a time window $\alpha_2dt$ much smaller than the time during which relaxation to stationarity takes place.
The optimization of the network parameters is then performed by minimizing the loss function
\begin{equation}
    L[\sigma_\theta](\theta) = \sum_t | \Delta_{\alpha_2} [Z]_t - \sigma^2_\theta(Z_t) |.
    \label{eq:loss_diffusion}
\end{equation}
Note that this loss function is insensitive to the sign of $\sigma_\theta$. This is not a problem since the stochastic increment $dW_t$ is symmetric under a change of sign. %process $Z_t$ includes the sign. 
For further details on the training procedure, we refer to the \ref{nn_ad_int}.
\section{The kinetic Ising model}
\label{sec:ising}
\subsection{The  model and its dynamics}
The Ising model is a paradigmatic model of statistical mechanics. 
It provides a qualitative description of the behavior of molecular magnetic dipoles in a metal.
The crystalline structure of the metal is modeled as a lattice of $N$ sites.
At each site $i=1,2,...,N $, a magnetic dipole is represented as a spin variable $s^i = \pm 1$.
The spins interact with each other according to the following energy functional (Hamiltonian)
\begin{equation}
    H(s) = 
    -\frac{1}{2}\sum_
    { \langle  ij\rangle}s^i s^j .
\end{equation}
Here, the notation $\langle  ij\rangle$ restricts the sites $i$ and $j$ in the sum to be nearest neighbors on the lattice. We consider a two-dimensional square lattice. 
This Hamiltonian presents a $\mathbbm{Z}_2$ symmetry since it is invariant under sign change of every spin variable $s^i \to -s^i$.
At thermal equilibrium at a given temperature $T$, each  spin configuration has a probability described by the Boltzmann distribution $\pi_{\rm B}(s)\propto e^{- H(s)/k_{\rm B}T}$, where $k_{\rm B}$ stands for the Boltzmann constant.
Given the magnetization 
\begin{equation}
\label{Mag_Ising}
    m = \frac{1}{N} 
    \sum_{i}s^i,
\end{equation}
the order parameter of the model is the expectation of the absolute value of $m$ in the Boltzmann distribution. 
The system undergoes a continuous transition from an ordered phase with finite magnetization at sufficiently low temperatures, to a disordered one with vanishing  magnetization. While in one dimension the model predicts a finite magnetization only at zero temperature, in two dimensions the critical temperature $T_c$ corresponding to the phase transition is finite.
Close to  $T_{\rm c}$, the value of the average magnetization $\bar m $ follows a power-law behavior
\begin{equation}
    \bar{m}\propto |T-T_{\rm c}|^\beta,
    \label{eq:beta_ising}
\end{equation}
where $\beta$ is a so-called critical exponent.

The Ising model discussed above does not possess inherent dynamics. In order to apply our ML method to this model we can endow it with 
Glauber dynamics using Metropolis-Hastings sampling, which is usually utilized for sampling the Boltzmann distribution of the model. 
Such a dynamic is defined by the single spin-flip  probabilities $P=  P(s^i\to-s^i)$, updating the spin variables in the lattice according to
\begin{equation}
    \label{eq:glauberMH}
    P(s^i_t \to -s^{i}_t)=
    \begin{cases}
    \exp(- \Delta E/k_{\rm B}T) \quad 
   & \text{if } \Delta E > 0
    \\
    1 &\text{if } \Delta E \leq 0
    \end{cases}
\end{equation}
where $ \Delta E = H(s^1_t,...,-s^{i}_t,..,s^N_t)-H(s^1_t,...,s^i_t,..,s^N_t)$ is the energy change associated with the transition.
For the completion of a single discrete time step $s_t\to s_{t+1}$, a single spin-flip is attempted $N$ times at a random site. 
%
%For initial value $m_0$ we consider the probability distribution $p$ of $m_t$ solving Eq. \eqref{eq:sde_ito}.
For such a {\it dynamical} Ising model, the (stochastically evolving) order parameter $m_t$
is defined as in Eq.~\eqref{Mag_Ising} for an evolving configuration $s_t$.
We choose each of the spins in the initial configuration to be up or down with equal probability, so that for large systems $m_0\approx 0$. 
For further detail about the model and its field theoretical representation, see \ref{ising_ft}.

\subsection{Neural network results}
\begin{figure}
    \centering
    \includegraphics[width=0.7\textwidth]{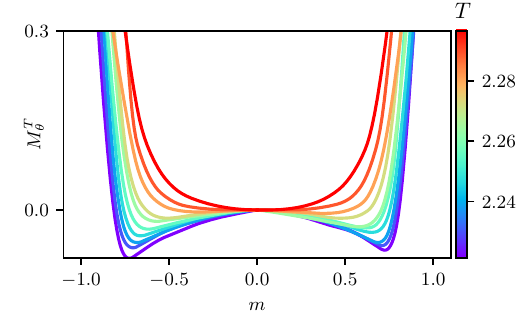}
    \caption{
    \textbf{Effective potential 
    $M_\theta^T$ of the Ising model.}
    The effective potential $M_\theta^T$ in Eq. \eqref{eq:eff_pot_ising} is determined by the integral of the drift $\mu_\theta$ with respect to the magnetization $m$.
    One can observe that for increasing  temperature this effective potential changes its functional form from having two distinct minima in the double-well potential to having only one minimum at $m=0$. 
       \label{fig:free_energy}
    }
\end{figure}
\begin{figure}
    \centering	
    \includegraphics[width=0.7\textwidth]{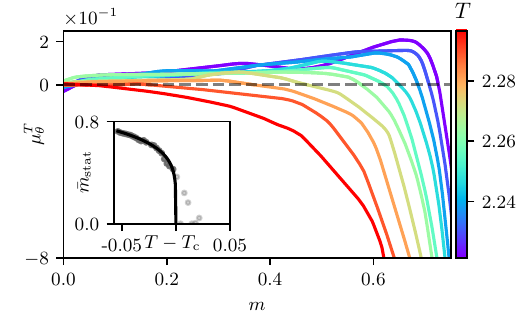}
    \caption{
    \textbf{Neural networks for the drift coefficients $\mu^T_\theta$ for the  
    Ising model.} 
    Due to the $\mathbb{Z}_2$ symmetry of the Hamiltonian, only the behavior for $m>0$ is shown. 
    Warmer tones correspond to higher temperatures. 
    The higher the temperature, the closer the non-trivial zero of $\mu^T_\theta$ corresponding to $\bar m_{\rm stat}$ is to zero. 
    Its value as a function of the distance to the critical temperature $T_{\rm c}$, obtained by minimizing the function $\epsilon(\tilde c_1, \tilde T_{\rm c},\tilde \beta)$ in Eq. (\ref{eq:epsilon}), is reported in the inset.}
    \label{fig:ising_beta_mu}
\end{figure}

Given a set of trajectories for $m_t$ at temperature $T$, we learn the corresponding drift term $\mu_\theta^{T}$ using the approach explained above and the loss function $L[\mu^T](\theta)$ in Eq.~\eqref{eq:loss_drift}. 
The drift term essentially acts as a directed force on the order parameter and it is thus natural to define an effective potential driving the motion of $m_t$ via the integral
\begin{equation}
    M_\theta^{T}(m) \equiv -\int ^m_0 dx \mu_\theta^{T} (x).
    \label{eq:eff_pot_ising}
\end{equation}
%UNCLEAR: which has the interpretation of the integral over space-time of the free energy of the model.
Our results reported in Fig.~\ref{fig:free_energy} show that upon increasing $T$ the effective potential undergoes a transition from a functional form exhibiting a double well  
to a single well potential. This fact is connected with the equilibrium Ising phase transition which  can be understood as follows. The stationary values of the expectation of the order parameter $\bar m_{\rm stat}$ correspond to the minima of the effective potential $M_\theta^T$, see Fig.~\ref{fig:free_energy}, and thus to zeroes of the drift coefficient, $\mu_\theta^T(\bar m_{\rm stat})=0$, cf. Fig. \ref{fig:ising_beta_mu}. Since the considered discrete-time dynamics samples the Boltzmann distribution at stationarity, one essentially has that the stationary values $\bar m_{\rm stat}$ should approximate the equilibrium order parameter $\bar{m}$, thus connecting the retrieved potential to the Ising transition. 

To benchmark the results from the trained networks $\mu^{T}_\theta$, we can thus extract the critical temperature $T_{\rm c}$ and the critical exponent $\beta$ of the order parameter and compare them with the known values for the Ising model. 
We fit the stationary magnetization $\bar m_{\rm stat}$ to the scaling form of Eq.~\eqref{eq:beta_ising} by minimizing the function 
\begin{equation}
    \label{eq:epsilon}
    \epsilon(\tilde  c_1,\tilde T_{\rm c},\tilde \beta ) = 
    \sum_{T}|
    \mu^{T}_\theta 
    (
    \tilde c_1 |T-\tilde T_{\rm c}|^{\tilde \beta}
    )|^2. 
\end{equation}
The positive function $\epsilon(\tilde c_1 , \tilde \beta, \tilde T_{\rm c})$ vanishes when $\tilde c_1 |T-\tilde T_{\rm c}|^{\tilde \beta} = \bar m_{\rm stat}$.
We consider the values of $c_1$, $T_{\rm c}$ and $\beta$ that minimize $\epsilon$ in Eq. \eqref{eq:epsilon}.
To find them, 
the zeroes of the drift coefficient $\mu_\theta^T$,
are computed using the exact derivatives 
via automatic differentiation. This is possible since % a feature which is allowed by the fact that 
we use differentiable neural networks. 
We find the following values:
$       \beta = 0.156 \pm 0.001$,       $ T_{\rm c} = 2.271 \pm 0.001$, and $
        c_1 = 1.076 \pm 0.002$. Note that the errors reported are only those related to the fit and do not consider finite-time and finite-size errors. 
For the Ising model, the analytical values are  $\beta=1/8$ and $T_{\rm c}=2/\ln(1+\sqrt{2}) \cong 2.269$ \cite{Onsager1944}. Our results are thus in good agreement with the exact values and show that the networks are able to provide a sound description of the critical behavior encoded in the data they are trained on.

Close to the critical point, the Ising model with Glauber dynamics is expected to fall in the
model A class according to the Halperin classification \cite{Halperin1977}. 
This is a pure relaxation model for a time dependent field in a double well potential, subject to uncorrelated white noise \cite{Debeye1965, Kawa1966,Coniglio1989,Satya1996}.
The latter feature is indeed reflected in our results on the learned diffusion coefficient $\sigma_\theta$, shown in 
Fig.~\ref{fig:mu_loss_ising}(b,c). There, we present $\sigma_\theta$ for $T=2.269$, which is in proximity to the critical temperature. As can be seen, the diffusion coefficient $\sigma_\theta$ is essentially constant when compared with the drift coefficient, entailing white noise in the dynamics of $m_t$.

We thus showed how the learned networks are able to encode  significant information about the statics, i.e., the order-disorder
phase transition (see Fig. \ref{fig:ising_beta_mu}) 
and the dynamics, i.e.,  the form of the noise, for the process under investigation, through a simple equation.
\section{The contact process}
\label{sec:contact}
\subsection{The  model}
We now apply our method to a paradigmatic nonequilibrium process, the so-called contact process \cite{Harris1974,Durrett1984}. It was introduced  
 to describe epidemic spreading in the absence of immunization. It is not defined via an energy function but solely via dynamical rules. 
The contact process shows a nonequilibrium continuous phase transition which belongs to the directed percolation universality class \cite{Broadbent1957, Geza2004,Domany1984,Hinrichsen1999,Liggett1985}.

Within the epidemic spreading interpretation of the model, each lattice site $i$ represents an individual which can either be found in the healthy state $s^i_t=0$ (inactive site) or in the infected state $s^i_t=1$ (active site). We consider here the case of a one-dimensional lattice. The dynamics occur in discrete time as follows: First, given the configuration $s_t$ at time $t$, we calculate the probability that each spin flips through the rules   

\begin{equation}
\begin{split}
\label{eq:cp_transitions}
    & P[0\rightarrow 1,n^i_t]\propto \kappa dt n^i_t / 2 \, ,\\
    & P[1\rightarrow 0,n^i_t]\propto \gamma dt \, .
    \end{split}
\end{equation}
Here, we introduced the healing rate $\gamma$, the infection rate $\kappa$ and  $n^i_t$ indicates the number of infected nearest neighbors of $i$. Then, according to the above probability, a spin is extracted, and the corresponding flip is performed. 
The order parameter 
is the number density of infected sites  
\begin{equation}\label{bar_rho}
    {\rho}_t = 
    \frac{1}{N}
        \sum_{i } s^i_t\, ,
\end{equation}
with $N$ being the total number of sites. 
We here consider $\rho_0=1$ as the initial value for the density (all sites infected).

From the dynamical rules in Eq.~\eqref{eq:cp_transitions}, one can see that the state with all healthy sites is a stationary state. In fact, this is a so-called absorbing state since it can be reached during the dynamics but it cannot be left. For any finite system, there is always a finite probability of hitting the absorbing state, which is the unique stationary state of the system. In the thermodynamic limit ($N\to\infty$) and for sufficiently large infection rates,  a phase with a finite density of infected sites, usually called  fluctuating  phase \cite{Hinrichsen1999, Liggett1985, Hinrichsen2000}, becomes stable. In finite systems, 
this phase eventually dies out and only appears within a meta-stable timescale. The absorbing phase and the fluctuating phase are separated by a continuous phase transition occurring at a finite critical value of the infection rate $\kappa_{\rm c}$, above which the system features a nonzero expectation of the stationary density $\bar{\rho}_{\rm stat}$. 
In proximity to the phase transition, the density follows a power-law behavior
\begin{equation}
    \Bar{\rho}_{\rm stat} \propto |\kappa-\kappa_{\rm c}|^\beta.
    \label{eq:beta_cp}
\end{equation}
In the following, 
we focus on a one-dimensional lattice made of $100$ sites and measure the infection rate in units of $\gamma$.
\subsection{Neural network results}

\begin{figure}
\centering
    \includegraphics[width=0.7\textwidth]{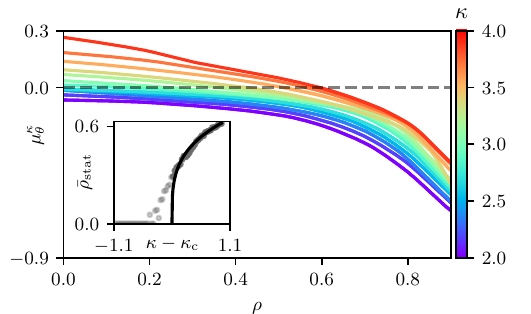}
    \caption{
    \textbf{Drifts $\mu_\theta^{\kappa}$ as a function of $\rho$, for different values of the infection rate.} 
    The zeroes of the drift term are associated with the stationary values of the stable stationary points 
    $\bar \rho_{\rm stat}$. 
    As the infection rate decreases, the stationary density collapses to $\rho = 0$. 
    From these values, 
    the critical exponent $\beta$ can be extracted in the proximity of the critical infection rate $\kappa_{\rm c}$, as shown in the inset. The drift terms in the active phase do not vanish when approaching $\rho=0$ as one may expect. As discussed in the main text, this is a consequence of the fact that trajectories in the active phase do not visit sufficiently often values below the stationary values  $\bar{\rho}_{\rm stat}$, which affects the training procedure.
    }
    \label{fig:cp_beta_mu}
\end{figure}

We start by discussing the results for the drift term of the contact process.
As for the kinetic Ising model, 
we train the network for many data sets of trajectories. 
For each data set at infection rate $\kappa$,
we train a model $\mu_\theta^{\kappa}$.
The results for learned drifts $\mu_\theta^{\kappa}$ are shown in Fig.~\ref{fig:cp_beta_mu}, for different
values of $\kappa$. Decreasing $\kappa$, the zero crossings $\mu^{\kappa}_\theta(\bar \rho_{\rm stat}) =0 $ occur at progressively smaller values of $\bar\rho_{\rm stat}$. In the inset, we illustrate how these  
can be used to extract 
the critical infection rate $\kappa_{\rm c}$ and the associated critical exponent $\beta$. As for the kinetic Ising model,
we can fit the density of infected sites to the power law of Eq.~\eqref{eq:beta_cp}
by minimizing  the function $\epsilon(\tilde c_1, \tilde \kappa_{\rm c}, \tilde \beta)$:
\begin{equation}
    \epsilon(\tilde  c_1,\tilde \kappa_{\rm c},\tilde \beta ) = 
    \sum_{\kappa}\left|\mu^{\kappa}_\theta \left(\tilde c_1 |\kappa-\tilde \kappa_{\rm c}|^{\tilde \beta}\right)\right|^2.
    \label{eq:epsilon_cp}
\end{equation}
The values $c_1, \kappa_{\rm c}$ and $\beta$ that
we find are $ \beta = 0.28 \pm 0.03$,
$ \kappa_{\rm c} = 3.062 \pm 0.003$.
These values should be compared with the values obtained by means of Monte Carlo or series expansion 
 $\kappa_c = 3.29785(8)$ \cite{Dickman1991, Dickman1998, jansen1993},  $\beta = 0.276486(8)$ \cite{Munoz1997,Jensen1999}.

Albeit this agreement, there is in fact a problem with the shape of the learned drifts $\mu_\theta^{\kappa}$, as shown in Fig.~\ref{fig:cp_beta_mu}. Given that the contact process features an absorbing state at density $\rho=0$, one should expect that the drift vanishes for this density. This is evidently not the case here. The reason lies in the fact that the physics actually influences the way in which training data can be produced. In our case, we train the network considering trajectories starting from the state with all sites infected. For such initial condition and being in the active phase, the density of infected sites will decrease with time until it reaches a (meta)stable finite value around which it will fluctuate. This implies that during the learning process values of the density smaller than the (meta)stable one, including the absorbing-state value $\rho=0$, are not visited sufficiently often. Therefore, it is not possible to appropriately learn the drift term below such values.

  \begin{figure}
  \centering
    \includegraphics[width=0.7\textwidth]{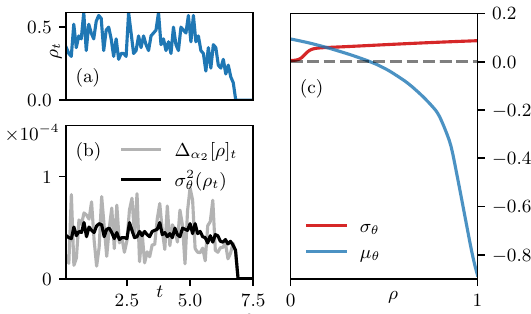}
    \caption{
    \textbf{Diffusion in the contact process. }
    (a) A sample trajectory $\rho_t$ for $\kappa = 3.36$. 
    Such trajectories $\rho_t$ are used to compute the derivative of the quadratic variation $\Delta_{\alpha_2}[\rho]_t$.
    This time derivative is used to train the network $\sigma_\theta$, via the loss in Eq.~\eqref{eq:loss_diffusion}.
    (b) The network takes as input the values of $\rho_t$ of a given trajectory and outputs the values $\sigma_\theta^2(\rho_t)$, which approximate the derivative of the quadratic variation $\Delta_{\alpha_2}[\rho]_t$, see Eq.~\eqref{eq:quad_var}. 
    Unlike in the Ising model, $\sigma_\theta$ is not a constant function of the order parameter $\rho_t$. The reason is that in the absorbing phase (where the order parameter is strictly zero) no fluctuations take place, as shown the same trajectory in panel (a).
 (c) Learned diffusion coefficient $\sigma_\theta$ as a function of the density $\rho$. We observe that $\sigma_\theta$ indeed goes to zero for vanishing order parameter $\rho$, signalling multiplicative noise. For comparison, we also provide the results for the corresponding drift $\mu_\theta$.}
    \label{fig:sigma_cp}
\end{figure}

In Fig.~\ref{fig:sigma_cp}, we report the results for the learned diffusion coefficient $\sigma_\theta(\rho)$, together with the network prediction for $\mu_\theta(\rho)$
and the time derivative of the quadratic variation $\Delta_{\alpha_2}[\rho]_t$, which the network learns (details on the network parameters are given in Table~\ref{fig:net_details}). 
We consider a value for the infection rate, $\kappa=3.36$, in the proximity of the critical point $\kappa_c$. 
In contrast to the Ising model, where both phases above and below the critical point are fluctuating, the presence of an absorbing phase dictates that the diffusion coefficient must vanish at zero density. This means that the noise must be multiplicative.
In fact, it can be proven that the diffusion coefficient is proportional to the square root of the density \cite{Hinrichsen2000,Janssen1981,Cardy1980}, which is a consequence of the central limit theorem and the fact that only active sites can contribute to fluctuations (for details we refer to \ref{cp_ft}). Both the learned diffusion coefficient $\sigma_\theta$ and the drift $\mu_\theta$ are not constant and approach zero for small $\rho$, see Fig. \ref{fig:sigma_cp}(c). They are not strictly zero at $\rho=0$ due to the above-discussed limitations of the learning procedure.

In Fig. \ref{fig:sigma_cp}(a) we show a selected trajectory, for which we display $\rho_t$, We see that $\sigma_\theta$ yields a time averaged value of the (coarse-grained) derivative of the quadratic variation $\Delta_{\alpha_2}[\rho]_t$ on which it was trained. Moreover, we also see that the learned noise vanishes as the system enters the absorbing state, i.e. $\rho_t=0$ [cf Fig. \ref{fig:sigma_cp}(b)].

\section{Conclusions}
\label{sec:concl}
We have shown how to encode a simple stochastic equation in an artificial neural network and applied this method to two 
paradigmatic models of statistical mechanics, both in and out of equilibrium. Both studied systems, the kinetic Ising model and the contact process, exhibit a continuous phase transition which also is captured by the network.  For both models we identified the critical point and retrieved the static critical exponent $\beta$. 

It is important to note that within the chosen approach the network does not learn the order parameter from raw configurations. Rather, it is fed with a one-dimensional average value of an order parameter (density or magnetization) and outputs the one-dimensional drift and diffusion coefficients for a given order parameter value.
The network thus learns one-dimensional quantities which simplifies the training process. In the case of the contact process, a multiplicative form of the noise is retrieved, while for the kinetic Ising model, the network learns a noise form that is approximately constant, i.e. independent of the value of the order parameter.

A natural future development would be to use the learned drift as a scaling function and to obtain all the critical exponents. This approach might also prove useful in classifying  universal behavior of different processes, as two models are expected to belong to the same class, not only if they share the same set of critical exponents, but also if they share the same scaling function. 
Another point for future exploration is to go beyond the inherently Markovian assumption in Eq.~\eqref{eq:sde_ito}, as the success of the results reported here, even under this assumption, could be attributed to the one-dimensional character of the training data. Future aims include the application of our approach to trajectories of open quantum processes and the utilization of machine learning methods that automatically infer the relevant order parameter \cite{Dietrich2022}.

\section*{Acknowledgments}
We acknowledge financial support from the Deutsche Forschungsgemeinschaft
(DFG, German Research Foundation) under Germany’s Excellence Strategy—EXC-Number 2064/1-Project Number 390727645 (the Tübingen Machine Learning Cluster of Excellence), EXC Number 2075-Project Number 390740016 (the Stuttgart Cluster of Excellence SimTech), under Project No. 449905436, and through the Research Unit FOR 5413/1, Grant No. 465199066. This project has also received funding from the European Union’s Horizon Europe research and innovation program under Grant Agreement No. 101046968 (BRISQ). F. Carollo is indebted to the Baden-W\"urttemberg Stiftung for the financial support of this research project by the Elite Programme for Postdocs.

\appendix
\section{Neural network details and integration of the learned stochastic equations}
\label{nn_ad_int}

Because of the different properties of the two models considered in the present work, the kinetic Ising model and the contact process, 
the employed networks and the 
hyper-parameters adopted to train them 
are slightly different. 
In the following, we specify the details of the networks and how the integration of the It\^o equation is performed. 
The code we use is available at \cite{Carnazza2023}.
\subsection{Neural network and training details}
\begin{table*}
    \centering
    \begin{tabular}[width=0.5\textwidth]{ |p{2cm}||p{2cm}|p{2cm}|p{2cm}|p{2.5cm}|  p{1cm}|p{1cm}|p{1cm}|}
     \hline
     \multicolumn{7}{|c|}{Network details} \\
     \hline
    Model &  Layers architecture  &  Learning rate (RMSprop optimizer) &Activation function & $n_{\rm train}$, $d_{\rm batch}$, $n_{\rm epochs}$&$\delta$,$\tau$& $\alpha_1,\alpha_2$  \\
     \hline
      $\mu_\theta$ Ising   &  $1\times 50 \times 50 \times 1$  &$0.3\times10^{-3}$&ReLU& 10, 100, 7000 &0.01, 1000&-  \\
     $\sigma_\theta$ Ising   &  $1\times 64 \times 1$  &$10^{-3}$&   Tanh (intra-layers), Sigmoid (output)& 10, 100, 5000 &-&1, 500\\
     $\mu_\theta$ Contact process &   $1\times 50 \times 1$  & $0.5\times10^{-3}$   &ReLU& 10, 100, 2000 &0.05, 100&-\\
     $\sigma_\theta$ Contact process   &  $1\times 64 \times 1$  &$10^{-3}$&    Tanh (intra-layers), Sigmoid (output)& 10, 100, 5000 &-&10,100\\

     \hline
    \end{tabular}
        \caption{ 
        \textbf{Details of the  networks' architecture and of the  training procedure.} 
        All the trained networks are fully connected feed-forward multi-perceptron networks \cite{Cybenko1989, Rosenbaltt1958}. 
        In the first column, the number of neurons $n_{i}$ in the $i$-th layer is reported, 
        as $n_1\times...\times n_{\rm out}$, with $i$ in $\{1,2,...,{\rm out}\}$.
        To train each network, the optimizer RMSprop algorithm in the PyTorch implementation is used, where only the learning rates were modified.
        The specific activation functions are also displayed.
        The quantity $n_{\rm train}$ refers to how many models are trained on each dataset, weighted averages of which are taken to compute the reference model.
        The dimension of the batch use to train is $d_{\rm batch}$, while $n_{\rm epochs}$ is the number of epochs . 
        The width of the interval from which $\mu$ is computed (cf. Fig. \ref{fig:mu_loss_ising}) is $\delta$, while the coarse graining of the discrete time is $\tau$.
        The constant $\alpha_1$ is used in computing the approximation for the quadratic variation 
        $[Z]_t = \sum_u (Z_{u+\alpha_1du} - Z_u )^2$.
        The constant $\alpha_2$ is instead using in computing its finite difference:
        $\sigma^2(Z_t) = ([Z]_{t+\alpha_2dt} - [Z]_t)/\alpha_2 dt$.
        }
    \label{fig:net_details}
\end{table*}
We model the drift $\mu_\theta$
as a fully connected feed-forward neural network.
The network is trained by employing back propagation methods to optimize 
the loss function \eqref{eq:loss_drift}
This optimization minimizes the distance between the 
function $\mu_\theta(x)$
and the drift coefficient $\mu$.
The back propagation lets us compute the gradients used in an optimization routine.
This routine requires as input a constant, namely, the learning rate, which amounts to the optimization step in the gradient descent algorithm.
The order of magnitude of the learning rate should be small enough to learn the data's essential details yet not too small to avoid learning the noise effects. Moreover, lower learning rates make the optimization procedure slower.
The learning rate we choose is thus a 
compromise between the optimization velocity and the accuracy of the results.
We optimize the network to learn $\Delta_1Z_t \tau$.  
The learned $\mu_\theta$ then has to be multiplied with $\tau$ to make it comparable with the training data. For the Ising model, the time scale $\tau$ is set to $1000$. For the contact process, it is $100$. 
Similarly, the  network $\sigma_\theta$ is a fully connected feed-forward network. As for $\mu_\theta$, the input and output dimensions are one-dimensional.
Both for $\mu_\theta$ and $\sigma_\theta$ the adopted optimizer is the 
PyTorch implementation of the RMSprop algorithm \cite{Paszke2019}.
The architecture and training details for the networks 
and $\sigma_\theta$ 
are reported in the Table \ref{fig:net_details}, both for the kinetic Ising model and the contact process.
For both of this processes the power law for the stationary values $\bar m_{\rm stat}$ and $\bar \rho_{\rm stat}$ only applies in the vicinity of the critical point, and only in the ordered and the active phase respectively.
For the Ising model, the sum in Eq.~\eqref{eq:epsilon} is computed for 15 values of the temperature equally spaced in between a minimum value
$T_{\rm min}= 2.2214$ and maximum value $T_{\rm max}=2.2759$.
Similarly, for the contact process, in the sum in Eq.~\eqref{eq:epsilon_cp},
we use 31 equally spaced values of the infection rate $\kappa$, from $\kappa_{\rm min }= 2.0$ as lowest value to highest value  $\kappa_{\rm max}=2.9831$.

To find the best critical values in Eqs. \eqref{eq:epsilon} and \eqref{eq:epsilon_cp}, we use the minimization library \cite{Feinman2021} that 
allows to compute exact gradients on the (differentiable) networks $\mu_\theta$ and $\sigma_\theta$ and minimize them numerically.

\subsection{Integration of the It\^o equation} 
\begin{figure*}
    \centering
    \includegraphics[width=0.49\linewidth]{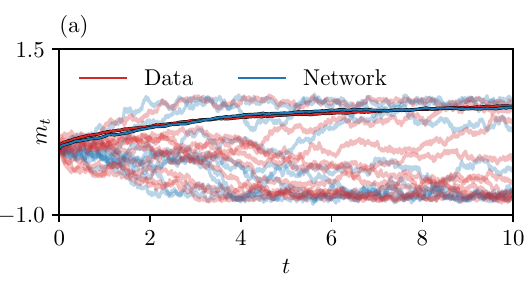}
    \includegraphics[width=0.49\linewidth]{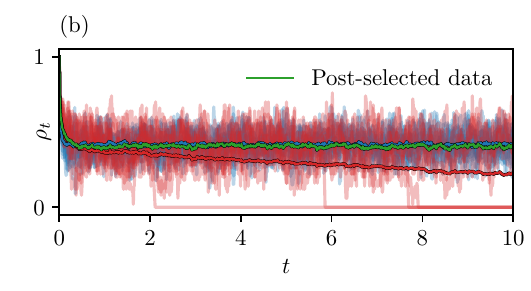}
    \caption{
    \textbf{Dynamical trajectories for the kinetic Ising model (a) and the contact process (b).}
    We report the ground truth trajectories (``Data", in red) as well as those generated by numerically integrating the learned It\^o Eq. \eqref{eq:nn_sde_ito} (``Network", in blue). 
    In panel (a), the time evolution of the magnetization of a two-dimensional lattice of size $N= 128 \times 128$ is shown for a temperature $T=2.2214$.
    Some samples for the ground truth trajectories 
    and the ones obtained from the network in lighter colors are displayed together with their averages over $100$ trajectories in darker ones.
    Analogously, in panel (b), the time evolution of the density of active sites $\rho_t$ on a one-dimensional chain of length $N = 100$ is shown for $\kappa = 3.36$, both for the ground truth and for the data generated by the network together with the respective order parameter estimated from $100$ trajectories. The green line (``Post-selected data") represents the order parameter computed only from trajectories which do not decay into the absorbing state $\rho=0$ (for the  
    Ising model, the order parameter is the trajectory average of the absolute value of the magnetization $m_t$ at time $t$).
     }
    \label{fig:trajectories}
\end{figure*}

In the present work, we extract an approximation to the 
drift coefficient $\mu$ and the time derivative of the quadratic variation $[Z]_t$ from the ground truth data.
It is interesting to numerically solve the learned It\^o Eq. \eqref{eq:nn_sde_ito} and compare the results with the ground truth $Z_t$. This can be readily done with the machine learning library Torchsde \cite{Kidger21b}, which we adopt here.
The numerical integration of It\^o equations requires small time steps to achieve convergence. The fictitious time scale $dt$ introduced to train the drift $\mu_\theta$ thus comes in handy for the integration. 
One needs to pay attention to the increments $W_{t+u}-W_t$ 
of the Wiener process in Eq. \eqref{eq:sde_ito}, which should be distributed with a probability $p$ following 
a normal distribution $\mathcal{N}(0,u)$ centered around zero and with variance $u$ such that $p(W_{t+u}-W_t)=\mathcal{N}(0,u)$.
The learned $\sigma_\theta$, 
which is trained without using the time scale $dt$, has thus to be divided by $\sqrt{dt}$ to make it comparable with the values obtained from the drift. 

The results obtained by integrating the learned It\^o Eq. (\ref{eq:nn_sde_ito}) present a similar behavior to the ground truth dynamics, see Fig. \ref{fig:trajectories}(b). 
For the contact process, our goal is to describe the dynamics up to the non-absorbing stationary state. For this reason, we 
restrict the training of the drift function $\mu_\theta$ to $\rho_t > \delta$ in the data set,
neglecting 
the region near the absorbing state ($\rho_t=0$). 
This allows us to consider short trajectories while retaining important information about how the active (non-absorbing) stationary state $\rho_{\rm stat}$ is reached. This implies 
that the learned drift function $\mu_\theta$ has only one stationary state, that is, a zero, in $\rho_{\rm stat}$, but not in $\rho_t=0$. 
When integrating the 
It\^o equation, no trajectory thus goes to the absorbing state, something that instead happens to the ground truth data. 
The average of the ground truth data (referred to as ``Data" in Fig. \ref{fig:trajectories})
thus slowly decreases towards zero,
while the average of only those trajectories in the ground truth data that do not end in 
the absorbing state exhibits a non-zero stationary state (indicated by ``Post-selected data" in Fig. \ref{fig:trajectories}). The latter agrees with the average obtained from integrating the learned It\^o equation (indicated by ``Network" in Fig. \ref{fig:trajectories}).

\section{Field-theoretic formulations}

\subsection{
Kinetic Ising model}
\label{ising_ft}

The field theory for the kinetic Ising model \cite{Halperin1977} is introduced 
by coarse-graining in space the originally discrete value of the spins $s^i$ 
in the critical regime. 
Averaging over some mesoscopic spatial volume, a real-valued spin density field $\psi$ is 
defined over a $(d+1)$-dimensional continuous space-time.
Specifically, the configurations $\Sigma_t$ have been coarse-grained so that at each point, a density $\psi \in \mathbbm{R}$ is defined. Then the stochastic time evolution for the density $\psi$ is provided by
\begin{equation}
    \partial_t \psi(\textbf{x},t)
    =
    -\frac{\delta \mathcal{F}[\psi]}
    {\delta \psi(\textbf{x},t)}
    +
    \eta(\textbf{x},t),
\end{equation}
with an effective potential functional of the form
\begin{equation}
    \mathcal{F}[\psi] = 
    \Gamma_0\int d^d\boldsymbol{x} 
    |\nabla \psi(\boldsymbol{x},t)|^2
    + u_0 \psi^2(\boldsymbol{x},t)
    + r_0 \psi^4(\boldsymbol{x},t)
    \label{eq:langevin_ising}
\end{equation}
and a Gaussian noise with correlations
\begin{equation}
    \langle 
    \eta(\boldsymbol{x}_1,t_1) 
    \eta(\boldsymbol{x}_2,t_2 )\rangle
    =
    2 \Gamma_0\delta(\boldsymbol{x}_1-\boldsymbol{x}_2)\delta(t_1-t_2).
\end{equation}
The diffusion constant $\Gamma_0$ and the coupling constants  
in \eqref{eq:langevin_ising} are functions of the model parameters. 

\subsection{Contact process}
\label{cp_ft}

For the field-theoretic formulation of the contact process, 
the average over some mesoscopic box in the lattice defines a coarse-grained density field $\rho(\textbf{x},t)$ (instead of taking the average over the whole lattice $\Bar{\rho}$). 
The Langevin equation for its time evolution
can be derived directly from the master equation of the contact process and reads \cite{Janssen1981,Hinrichsen2000}
\begin{equation}
    \partial_t \rho(\textbf{x},t)=
    D \nabla^2 \rho(\textbf{x},t)
    + \iota \rho (\textbf{x},t)
    - \lambda \rho^2(\textbf{x},t)
  +  \zeta(\textbf{x},t).
  \label{eq:langevin_cp}
\end{equation}
The noise $\zeta(\textbf{x}, t)$ exhibits a multiplicative form
\begin{equation}
    \langle 
    \zeta(\boldsymbol{x}_1,t_1) 
    \zeta(\boldsymbol{x}_2,t_2 )\rangle
    =
    \Gamma   \rho(\boldsymbol{x}_1,t_1) \delta(\boldsymbol{x}_1-\boldsymbol{x}_2)\delta(t_1-t_2).
\end{equation}
$D$ is the diffusive constant, while the coupling constants $\iota$, $\lambda$, and $\Gamma$ are functions of the lattice details and of the infection rate.
The occurrence of a term proportional to $\rho$ and one proportional to $\rho^2$ 
in Eq. \eqref{eq:langevin_cp} 
can be explained heuristically from the mean-field treatment of the transition rates in Eq. \eqref{eq:cp_transitions}. The number of sites becoming inactive at time $t$ is $\sum_{i} s^i_t\propto \rho_t$.
Instead, the number of sites becoming active is given by the number of inactive sites next to an active site that can be thus be infected.
This number is given by $\sum_{i} (s^i_t-s^{i+1}_t)^2 = 2\sum_{i }s^i_t - 2\sum_{i }s^i_t s^{i+1}_t\propto 2\rho_t -2\rho^2_t$.
In the mean-field treatment, the master equation thus reads $\partial_t \rho_t = (\kappa-1) \rho_t - \kappa \rho^2_t$.
The form of the noise
proportional to $ \sqrt{\rho}$,
can be justified by observing that 
only active sites contribute to the density fluctuations. 
To see this, let $N$ be the total number of sites, and let $n$ be the number of active sites at time $t$. 
The density of active sites at time $t$ is thus $\rho_t=n/N$. 
Now, let $X_i$
be the number of active sites at time $t'>t$ whose infection can be traced back to the 
$i$-th active site at time $t$.
Notice that the sequence 
$\{X_1,...,X_n\}$ is formed by independent identically distributed random variables, and $\rho_{t'}= 1/N \sum_i X_i$.
Their expectation value and variance will thus be independent of the site $i$:
$\mathbb{E}[X_i]=\nu$, 
${\rm Var}[X_i]=\zeta^2$ for some real number $\nu$ and $\zeta$.
The relation between $\nu$, $\zeta^2$
and the sample average 
$\bar X_n = 1/n \sum_i{X_i}\equiv \rho_{t'}/\rho_{t}$ is described by the central limit theorem.
This theorem states that for large $n$, the probability distribution $p$ of the random variable $\sqrt{n}(\bar X_n -\nu)$
converges to a normal distribution centered around zero and with variance $\zeta^2$, $\mathcal{N}(0,\zeta^2)$:
\begin{equation}
    p\left(\sqrt{n}( \bar X_n -\nu )\right) \mathcal{N}(0,\zeta^2).
\end{equation}
Note that for $\rho_t$ to be finite also $N$ must be large. Substituting $\sqrt{n}\to N \rho_t $ and $\bar X_n \to \rho_{t'}/\rho_t$, one obtains
\begin{equation}
    p\left(\sqrt{N}( \rho_{t'} - \rho_t\nu )\right) \to \sqrt{\rho_t}\mathcal{N}\mathcal(0,\zeta^2),
\end{equation}
which means that the expectation value of $\rho_{t'}$ is $\rho_t \nu$, and its variance $\rho_t \zeta^2$.

\section*{References}
\bibliographystyle{iopart-num}
\bibliography{bib}

\end{document}